\documentclass[10pt,aps,pre,showpacs,twocolumn]{revtex4}

\usepackage[utf8]{inputenc}
\usepackage{amsmath}
\usepackage{mathptmx}
\usepackage{graphicx}
\usepackage{color}
\usepackage{amssymb}

\begin{document}

\title{Statistics of Vector Manakov Rogue Waves}
\author{A. Man\v ci\' c$^1$, F. Baronio$^2$, Lj. Had\v zievski$^3$, S. Wabnitz$^{2,4}$, A. Maluckov*$^{3}$}
\affiliation{$^1$Faculty of Sciencies and Mathematics, University
of Ni\v s, POB 244, Ni\v s, Serbia}

\affiliation{$^2$INO-CNR and Dipartimento di Ingegneria dell'Informazione, Universit\`a di Brescia, Via Branze 38, 25123 Brescia, Italy}

\affiliation{$^3$P$^*$ group, Vin\v ca Institute of Nuclear
Sciences, University of Belgrade, POB 522, Belgrade, Serbia}

\email{sandram@vin.bg.ac.rs}

\affiliation{$^4$Novosibirsk State University, Novosibirsk 630090,
Russia}

\begin{abstract}
We present a statistical analysis based on the height and return
time probabilities of high amplitude wave events in both focusing
and defocusing Manakov systems. We find that analytical
rational/semirational solutions, associated with extreme, rogue
wave (RW) structures, are the leading high amplitude events in
this system. We define the thresholds for classifying an extreme
wave event as a RW. Our results indicate that there is a strong
relation between the type of the RW and the mechanism which is
responsible for its creation. Initially, high amplitude events
originate from modulation instability. Upon subsequent evolution,
the interaction among these events prevails as the mechanism for
RW creation. We suggest a new strategy for confirming the basic
properties of different extreme events. This involves the
definition of proper statistical measures at each stage of the RW
dynamics. Our results point to the need for defining new criteria for identifying RW events.

\end{abstract}

\pacs{05.45.Yv, 02.30.Ik, 42.65.Tg}

\maketitle

\section{Introduction}

The emergence, dynamics and prediction of rogue waves (RW), also
referred to as freak waves or extreme events, has been in the
focus of interest in diverse fields of science (oceanography,
physics of fluids, optics, matter waves physics, sociology,
bio-sciences,...) over the last fifteen years \cite{1,2,3,3a}.
However, there are still more open questions than answers
concerning the definition, genesis, dynamics, predictability and
controllability of RW phenomena \cite{4,5}. This
RW debate has stimulated the comparison of predictions and
observations among distinct topical areas, in particular between
optics and hydrodynamics \cite{rep,a2}.

Peregrine solitons \cite{c1} and Akhmediev
breathers \cite{c2} are well-known RW candidates: they represent
solutions of the scalar one-dimensional self-focusing nonlinear
Schrodinger equation (NLSE); the Peregrine solitons with the
property of being localized in both the transverse and evolution
coordinates, the Akhmediev breathers being periodic in the transverse
coordinate and localized in the evolution dimension. The Peregrine type
solitons are unique also in a mathematical sense, since they are
written in terms of rational functions of coordinates, in contrast
to most of the other known solutions of the NLSE, which are purely
exponential.
Recent experiments have
provided a path for generating Peregrine solitons in optical
fibers with standard telecommunication equipment \cite{4}, as well
as in water-wave tanks \cite{5a,5b}. To the contrary, in the
scalar case the defocusing nonlinear regime does not allow for RW
solutions, even of a dark nature.

Recently, progress has been made by extending the search for RW
solutions to coupled-wave systems. Indeed, numerous physical
phenomena require to model waves with two, or more components, in
order to account for different modes, frequencies, or
polarizations. In those cases, the focusing regime is not a
prerequisite for the existence of RW solutions. When compared with
scalar dynamical systems, vector systems may allow for energy
transfer between the coupled waves, which may yield new families
of vector RW solutions (bright-bright, bright-dark type), with
relatively complex dynamics. Such types of RWs have recently been
found as solutions of, e.g., the focusing vector NLSE
\cite{5,6,7,8}, the three-wave resonant interaction equations
\cite{9,10}, the coupled Hirota equations \cite{11}, and the
long-wave-short-wave resonance \cite{12}. It is crucial to add
that new RW families can be created in the defocusing nonlinear
regime too. This was shown theoretically and experimentally in
\cite{6,13,b1,b2,b3}: its authors proved that, in the defocusing
regime of the Manakov system, the range of existence of rational
solutions of different types (bright-dark, dark-dark), which are
the most serious candidates for RW, overlaps with the region of
baseband modulation instability (MI). Moreover, it was
demonstrated that MI is a necessary but not sufficient condition
for the existence of RWs. It is generally
recognized that MI is one of the mechanisms for the RW generation,
and recent observations of higher-order MI on the water surface have
been reported \cite{5c}.

However, a basic question arises regarding the statistical
description of high amplitude events in the course of nonlinear
wave propagation. It should be considered that under realistic
circumstances the propagation medium exhibits fluctuations of its
parameters, hence of the background continuous wave (CW)
solutions. To describe both bright and dark structures on a
background, the term high amplitude wave is used in the sense that
it denotes either high amplitude peaks or dips on a background. In
addition, it is important to develop a global understanding of RW
emergence in a turbulent environment, which connects with the
broad topic of wave turbulence in integrable systems \cite{14}. In
this respect, we may distinguish two ways of seeding MIs. The
first mechanism is associated with noise-driven MI. It refers to
the amplification of initial noise superposed on a plane wave
solution, which leads to spontaneous pattern formation from
stochastic input wave fluctuations. The second mechanism is that
of coherently driven MI, which refers to the preferential
amplification of a specific perturbation (thus leading to a
particular breather solution) with respect to broadband noise. It
was shown that breather wave dynamics is subject to competitive
interactions of the two types of seeding of MIs \cite{14}.
Nevertheless, a complete physical picture of these various
phenomena is still lacking.

Moreover, outside the context of discrete systems and numerical
studies of supercontinuum generation \cite{new1}, the statistical
analysis has not yet found a leading role in the studies of RWs,
although RWs are statistically determined entities. In our
research, we shall provide a new insight into the origin and
dynamics of multiparametric vector RW solutions, by adopting a
statistical approach. A similar study has been very recently applied to characterize vector RW generation in highly birefringent optical fibers \cite{new}.
In that case, a key role in the RW generation mechanism is played by the presence of group velocity walk-off between the two polarization components,
and third order dispersion.

In this paper we statistically investigate the behavior of high
amplitude events in the integrable Manakov system. For this end,
we numerically model the (light/matter) wave propagation in the
nonlinear media (photonic/Bose-Einstein condensate), in the
simplest case of a two component system. Physically this
corresponds to the case of two orthogonal polarization states of
light, or two different atomic states in BEC \cite{bec16}. The
initial conditions of the wave system represent a crucial issue in
our study. In order to simulate fluctuations in the properties of
a real system, we shall consider the injection of plane waves with
additive white noise in the system. The long time numerical
simulations will be performed by means of the pseudo-spectral
Fourier method, in order to obtain a proper statistical ensemble
of high amplitude events. Note that the term time will be used as
a synonym of the propagation length in the following. A brief
description of applied numerical and statistical methods is
presented in Section 2. The results and their interpretation with
respect to different types of RW candidates, different mechanisms
of high amplitude events creation and their statistical and
dynamical properties are considered in details in Section 3.
All results lead us to conclude that new criteria
for identifying high amplitude events are necessary. Conclusions
and comments are given in Section 4.

\section{The model equations}

The vector nonlinear Schr\"{o}dinger equations, i.e., the Manakov
system, can be written in dimensionless form as

\begin{eqnarray}
i\frac{\partial u^{(1)}}{\partial z} +\frac{\partial^2
u^{(1)}}{\partial t^2}
-2s(|u^{(1)}|^2+|u^{(2)}|^2)u^{(1)}&=&0\nonumber\\
i\frac{\partial u^{(2)}}{\partial z} +\frac{\partial^2
u^{(2)}}{\partial t^2} -2s(|u^{(1)}|^2+|u^{(2)}|^2)u^{(2)}&=&0
,\label{eq1}
\end{eqnarray}

\noindent where $u^{(1)}(t,z)$ and $u^{(2)}(t,z)$ represent the
wave envelopes, $z$ is the evolution variable, and $t$ is a second
independent variable. The meaning of variables depends on the
particular applicative context (fluid dynamics, plasma physics,
BEC, nonlinear optics, finance). The parameter $s=-1$ refers to
the focusing (or anomalous dispersion) regime, while $s=1$ refers
to the defocusing (or normal dispersion) regime of wave
propagation in the nonlinear medium. Model equation (\ref{eq1}) is
fully integrable, and it can be solved by applying the Darboux
dressing technique \cite{6,13}. Being focused on high amplitude
events, we mention briefly the rational or semirational localized
solutions of Eq. (\ref{eq1}), which are considered as one of the
most promising candidates for RW events in the literature
\cite{6,13}.

\begin{figure}
\includegraphics [width=6cm]{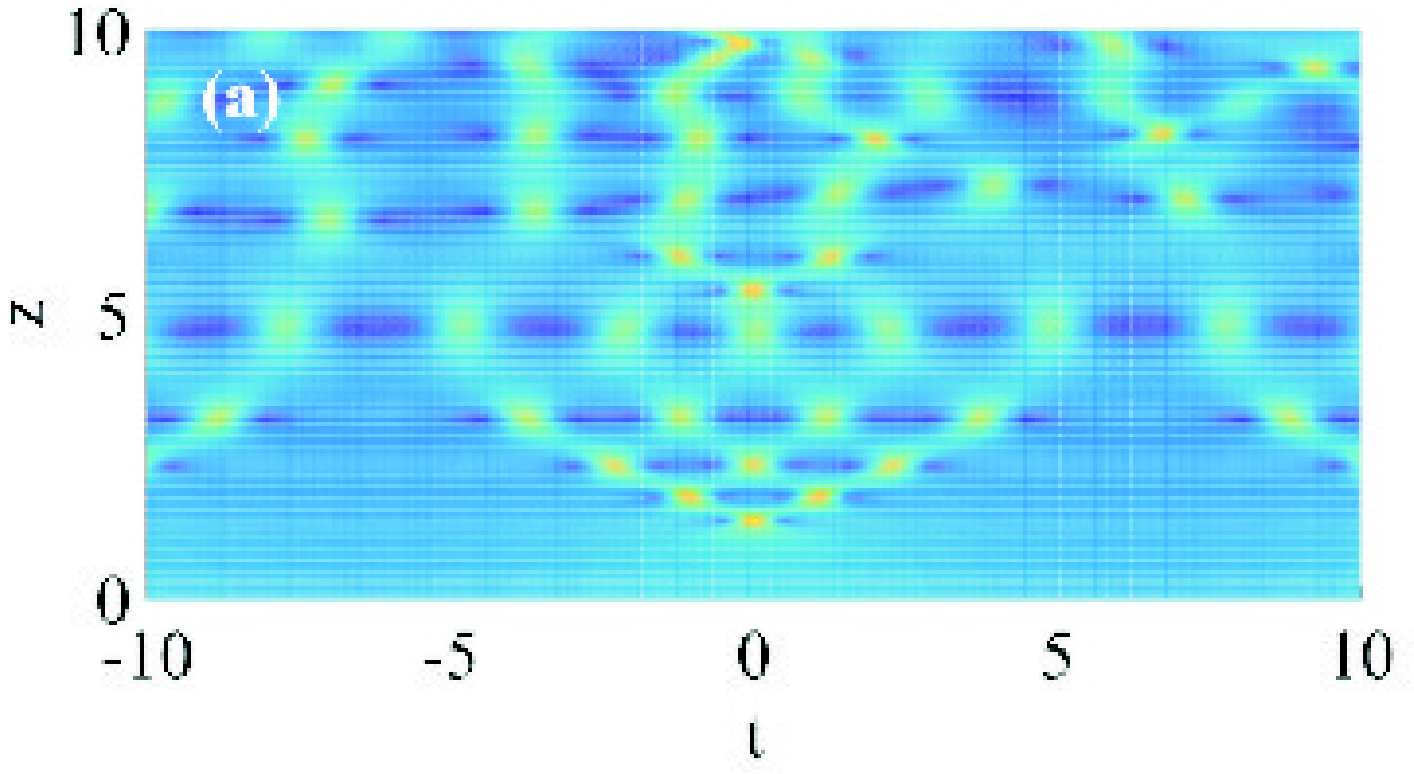}
\includegraphics [width=6cm]{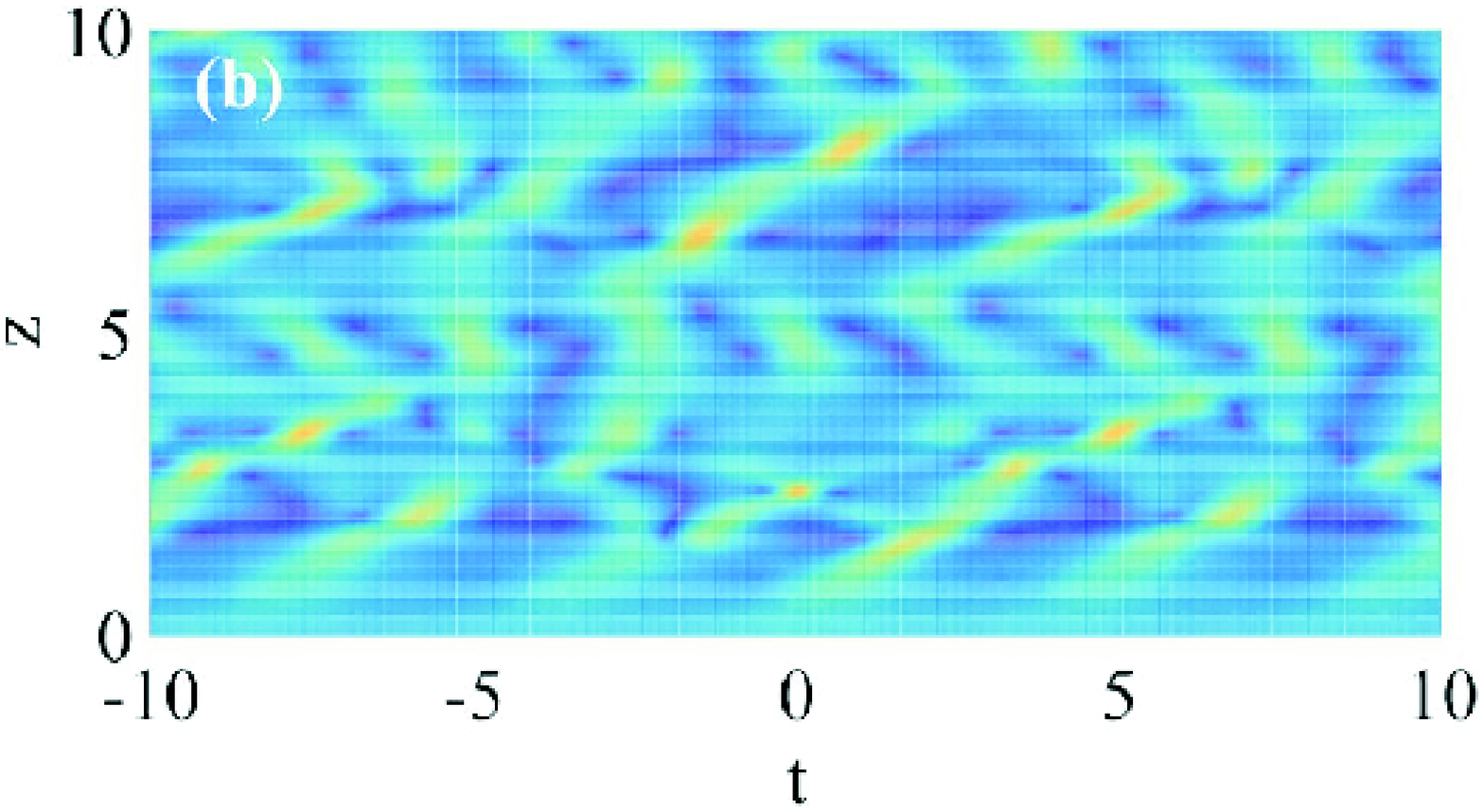}
\includegraphics [width=6.2cm]{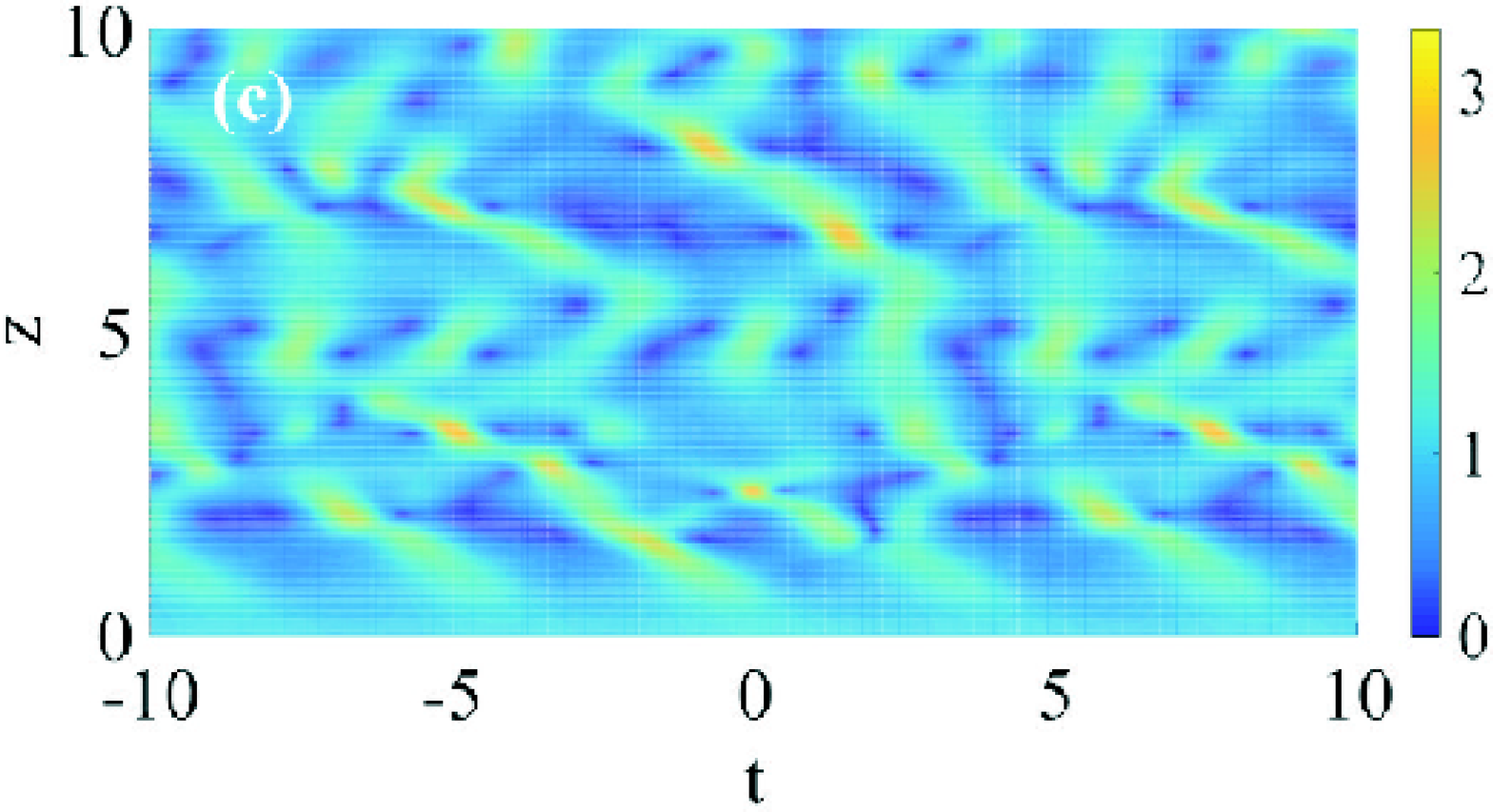}
\caption{Localized patterns in the focusing Manakov system for
$a_1(0)=a_2(0)=1$ and $q_1=q_2=0,\,q=0$ ($|u^{(1)}(t,z)|$) (a),
$q_2=-q_1=1, q=1$ (plots of two components $|u^{(1)}(t,z)|$ and
$|u^{(2)}(t,z)|$ are shown separately) (b) and (c).
The absolute values of the corresponding
amplitudes are shown in the box (i. e. the colorbar) in the last
plot, and it is mutual for all plots in the figure. Initial
small uniform random and periodic perturbations are added to the
plane wave background which is amplitude modulated by a super
Gaussian.} \label{fig1}
\end{figure}

In the focussing case, such rational solutions can be expressed in
the form of different bright-dark breather composites \cite{6}: e.g., a
boomeron-type soliton with a time-dependent velocity, a
breather-like wave resulting from the interference between the
dark and bright contributions, and more complex structures resulting from the merging of
Peregrine and breather solutions. The last case provides the evidence
of an attractive interaction between the dark-bright wave and the
Peregrine soliton solutions. In Fig. \ref{fig1} we present numerically
obtained examples of localized wave structures in the focusing case,
which are initialized by a plane wave in the form
\begin{equation}
u_0^{(j)}=a_j e^{i(q_j t-\nu_j z)},\quad
\nu_j=q_j^2+2(a_1^2+a_2^2),\quad j=1,2,\label{eq5}
\end{equation}
with simultaneously added small periodic and random perturbations.
The $a_j$ parameters represent the initial amplitudes of component
waves in the system, while $q_j$ are the initial phases. The
difference of phase factors $q_1-q_2=2q$ will be used to present
our numerical results in the next sections.

On the other hand, in the defocusing case the
rational/semirational solutions were explicitly derived in
\cite{13}. They can be generated both analytically and numerically
by starting from a plane wave solution (\ref{eq5}). It was
analytically shown \cite{13} that the region of rational wave
existence, which is related to the domain of RW existence, is
determined by the following expression

\begin{equation}
(a_1^2+a_2^2)^3-12(a_1^4-7a_1^2
a_2^2+a_2^4)q^2+48(a_1^2+a_2^2)q^4-64q^6>0.\label{eq6}
\end{equation}

\begin{figure}
\includegraphics [width=6cm]{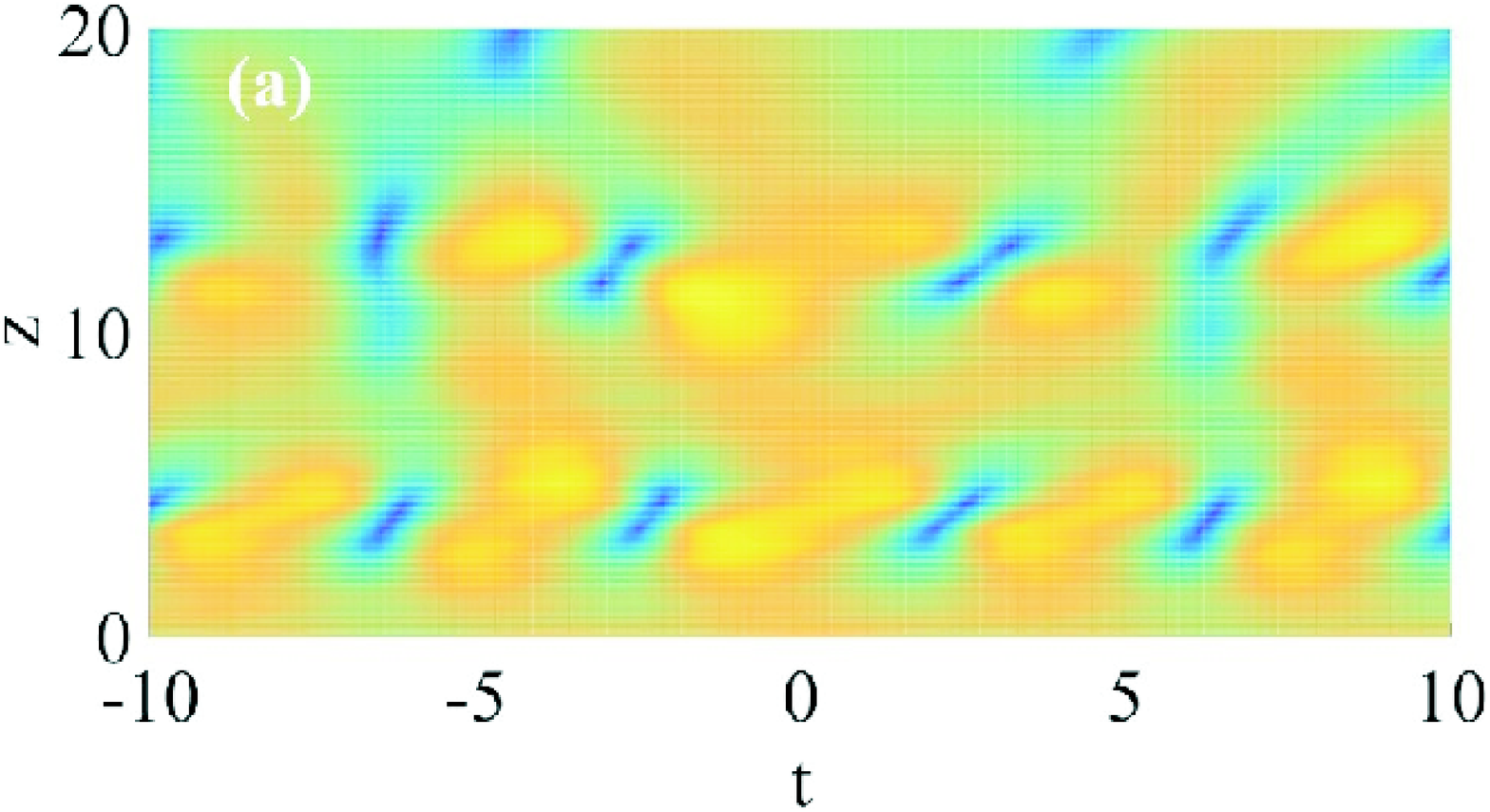}
\includegraphics [width=6cm]{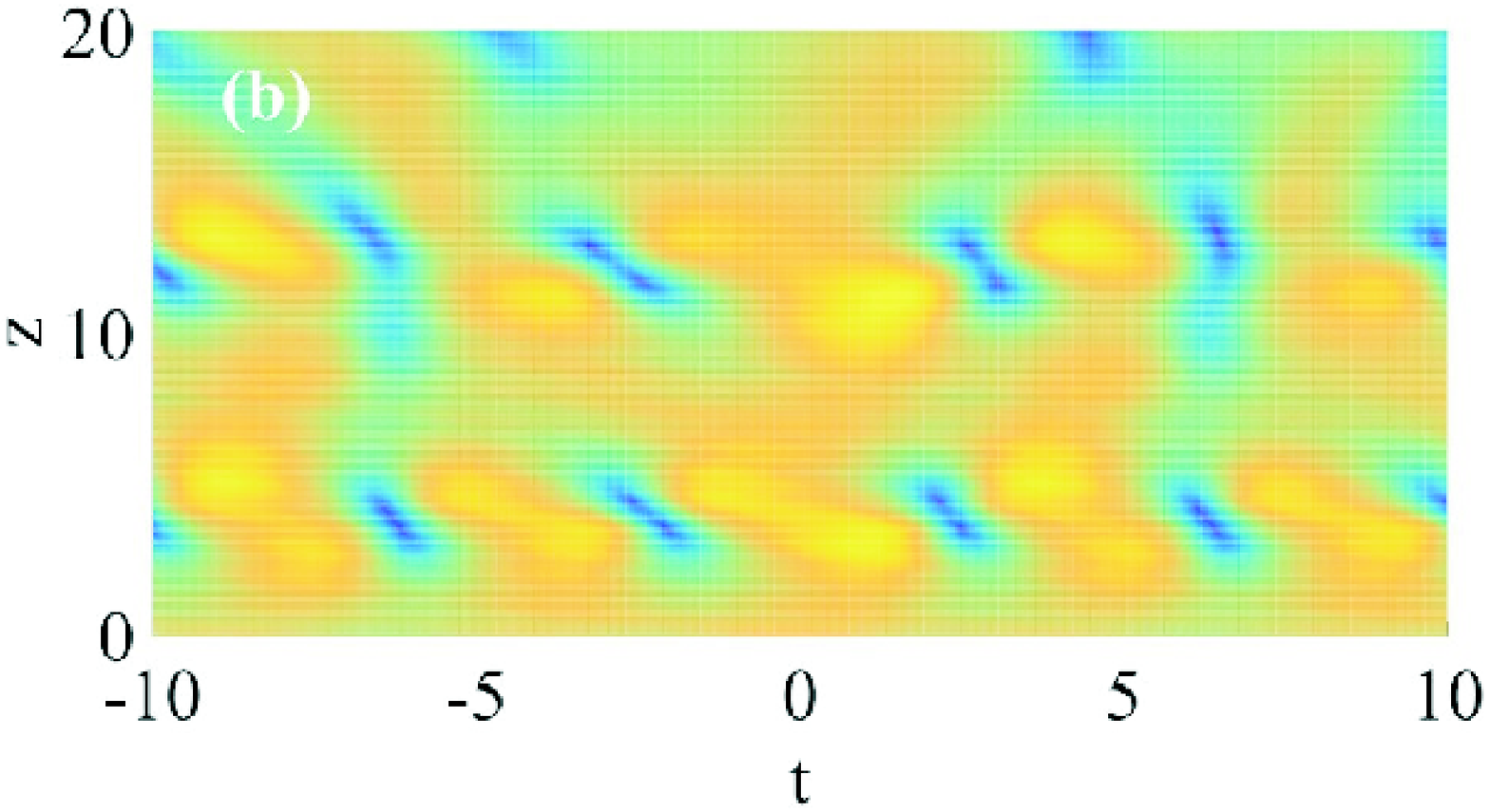}
\includegraphics [width=6.5cm]{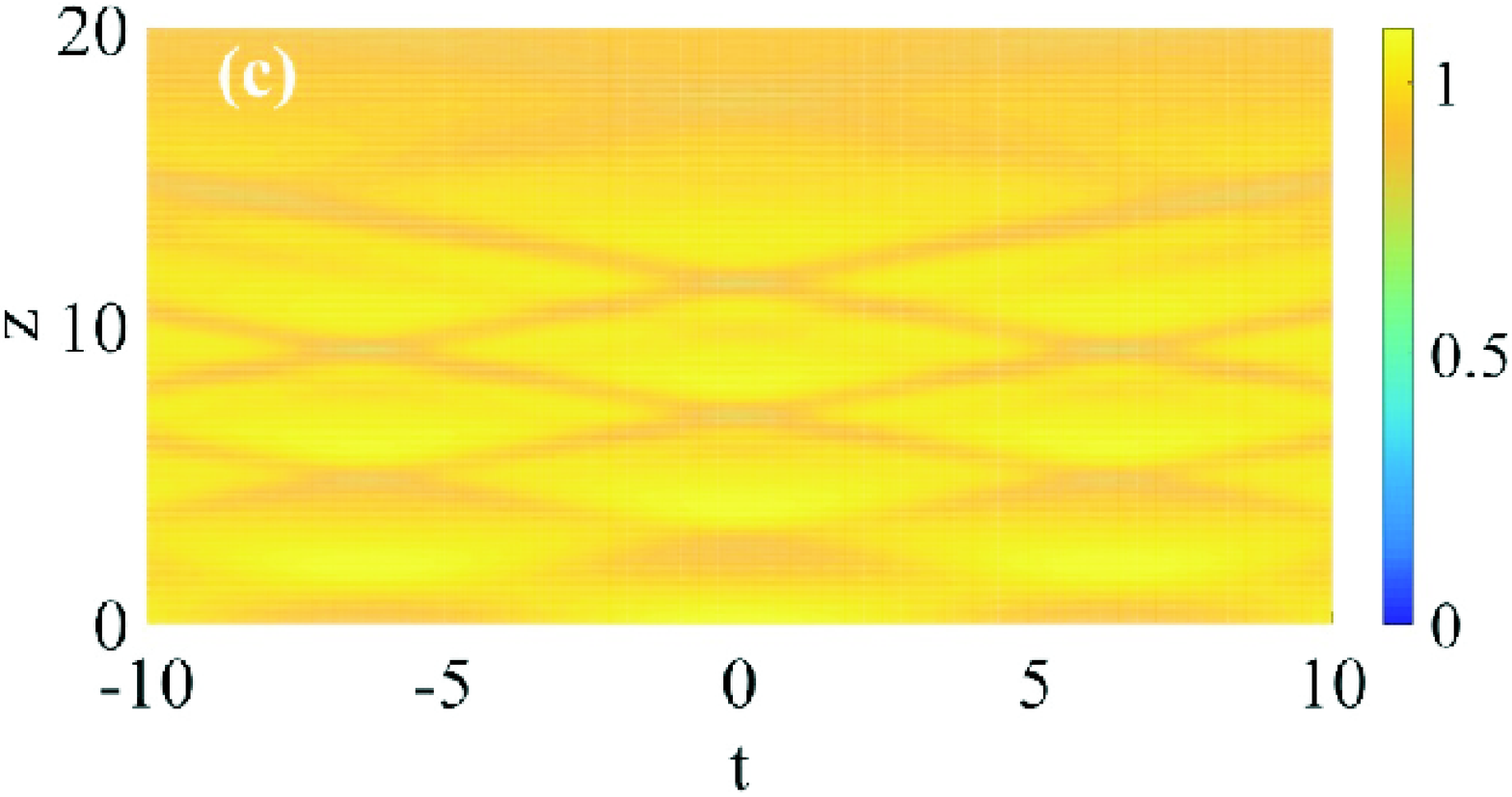}
\caption{Localized patterns ($|u^{(1)}(t,z)|$ and
($|u^{(2)}(t,z)|$) in the defocusing Manakov system for
$a_1(0)=a_2(0)=0.8$, $q_2=-q_1=1,\, q=1$ (a,b) (region of
existence of rational solitons). In plot (c) the case outside the
region of the rational RW existence reported in \cite{13} is
presented, with $a_1=a_2=1,\,q=0$ ($|u^{(1)}(t,z)|$).
The absolute values of the corresponding
amplitudes are shown in the box (i. e. the colorbar) in the last
plot, and it is mutual for all plots in the figure. Initially
small cosine and uniform random perturbations are added to the
plane wave background, amplitude modulated by a super Gaussian.}
\label{fig2}
\end{figure}

In particular, the inequality (\ref{eq6}) implies that the background
amplitudes have to be sufficiently large, for a fixed $q$, in order to allow
for the rational wave formation, see Fig. \ref{fig2}. Here, we prefer
not to use the term RW for high amplitude rational solutions, since
an unique definition of RWs does not exist. Indeed, the findings presented in
the following will sustain our terminology. Examples of these rational/semi-rational solutions of the
defocusing nonlinear Manakov system are shown in Fig. \ref{fig2}
(a,b). By adding to the finite background small regular
(periodic) and random perturbations in the parameter regimes
associated with the presence of MI, we confirmed the analytical predictions and
previous numerical results from the literature, \cite{13}. The preparation
of initial conditions included the presence of a super-Gaussian
amplitude modulation of the background. This was done in order to ensure
conditions that would isolate the MI mechanism from possible presence of numerical artifacts
(e.g., boundary reflections).

The next step was to prepare initial conditions that can ensure
the generation of a huge ensemble of localized, high amplitude events, which is
necessary for the statistical analysis.  We analyzed the results of
numerical simulations with different initial conditions, namely, a plane
wave (uniform background) with random perturbations (white noise,
Gaussian noise), with a small periodic (coherent) perturbation, and with a combination of
both small random and periodic perturbations. In all cases, qualitatively the
the same behavior was obtained. Therefore, we
decided to perform numerical simulations by injecting a noise
seeded plane-wave field into the model equations, (\ref{eq1}). We
applied the standard split-step numerical procedure for solving
the evolution equations, \cite{16}. In order to obtain
a qualitative confirmation of our numerical findings, we applied, in
parallel, the sympletic variants of the split-step method: SABA2
and SBAB2 algorithms \cite{sympl}. Qualitatively, the same results and
conclusions were obtained.

Amplitude noise is numerically modeled as a uniform random
process with zero mean. In order to have sufficient data for the
statistical analysis, the long term evolution of the field was numerically simulated. The optimal width of the calculation window was
estimated in each particular case by repeated numerical tests.

\section{Statistics of the Manakov rogue waves}

\begin{figure}[h]
\centering
\includegraphics[width=6cm]{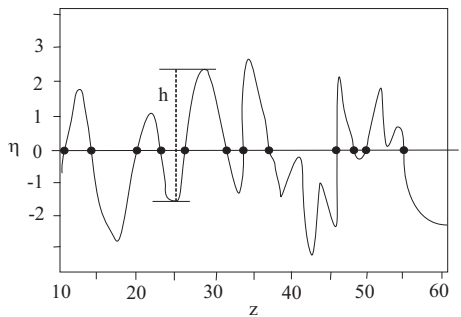}
\caption{The schematic illustration of determination of the wave
height. Quantity $\eta(z)=|u(z)|-|u(0)|$ is the wave amplitude
elevation \cite{rep}.} \label{skica}
\end{figure}

The purpose of this study is the statistical analysis of the
emergent peaks (dips) in the numerical solutions of the Manakov
system. Such extreme amplitude wave events are usually referred to
as RWs, whenever the significant height criterion is satisfied
\cite{17,statistics}. Here, the difference between the maximum
value of the finite background elevation in between two
zero-crossings and the minimum value of the background elevation
in the adjacent (next or previous) zero crossing interval is
called the wave height (Fig. \ref{skica}). In
scalar models of water-wave propagation, the significant height
$h_s$ is defined traditionally as the average height of one-third
of the highest waves in the height distribution, and the RW
threshold is estimated to be $h_{th}\ge 2.2h_s$ (also, in the
literature on ocean rogue waves, waves with height bigger than
$2h_s$  qualify to be in this category \cite{Kharif}.

%%%%%%%%%%%%%%%%%%%%%%%%%%%%%%%%%%%%%%%%%%%%%%%%%%%%%%%%%
%%%%%%%%%%%%%%%%%%%%%%%%%%%%%%%%%%%%%%%%%%%%%%%%%%%%%%%

In the preparatory phase of our study, we searched for proper RW
classifiers. Recently, a two-dimensional (2D) equivalent of the
significant wave height was defined as a classifier in vector
models (\cite{18}). In the framework of the complex RW patterns that are observed in our
model, this does not seem as an appropriate criterion to declare
that an event is of the RW type. Defining a new, proper
classifier(s) remains a challenge for future studies. Here, the
significant height criterion is slightly modified: we introduce a vector ($\hat{h}_s)$, where each
of its components measures the significant height of
the respective field component $(h^{(j)}_s,\, j=1,2$)

\begin{equation}
\hat{h}_s=(h^{(1)}_s \,h^{(2)}_s )^T.\label{eq6a}
\end{equation}

In this expression, the abbreviation $T$ indicates the transpose
operation. Thus, the height threshold, ($\hat{h}_{th})$, is a
vector quantity consisting of the height thresholds with respect
to two spinor components, $(h^{(j)}_{th},\, j=1,2$). Finally, if
at least a height of one of the components reaches the
corresponding threshold height, the event is declared as a RW. Let
us note that the threshold criterion for each particular component
is the same as the usual one for the one-dimensional case:
$h^{(j)}_{th}=2.2 \,h^{(j)}_s,\, j=1,2$. However, the proper
definition of the height criterion for a RW in multi-component
system remains still an open issue. For the sake of simplicity,
the vector abbreviations for significant height and threshold
height will be omitted in the following
($\hat{h}_s=h_s,\hat{h}_{th}=h_{th})$.

We calculated different statistical measures which have been developed in the
literature on extreme events, and considered their relevance for
expressing the dynamical properties of high amplitude events in the
Manakov system. It was shown that the most adequate statistical measure for
our system is that based on the height and
return time, namely, the probability density of the wave height $P_h$, or height probability density (HPD)
\cite{nash,new}, coupled with the probability distribution of the return time among to successive RW events, i.e.,
$P_r$, \cite{20}.

In the following, we will discuss the shape of the $P_h$ curves
(associated with the corresponding moments) as a function of
$h_s$, along with the probability of RW occurrence $P_{ee}$, which
can be derived from $P_h$. The tails of the HPD are related to the
presence of extreme events. The probability of RW occurrence is
defined as $P_{ee}=P_h (h>h_{th}=2.2h_s)$ (with respect to both
vector components), and it is obtained by integration of the
normalized $P_h$ from $h=h_{th}$ up to infinity.

For a deeper insight into the time statistics of RWs, the
probability distribution of the return time (time is a synonym of
propagation length/duration), $P_r$ of these (vectorial) events
was also calculated. The return time $r$ is defined as the time
interval between the appearance at given position of two
successive events with amplitudes above a certain predetermined
height threshold $h_r$. Details on the calculation of the return
time probability distribution are given in \cite{20}. Briefly, the
return time is registered as the time interval between two
successive events with a height (i.e., heights of both field
components) above a certain threshold value, which appears at the
same given lattice location. We follow this procedure repeatedly
up to the end of our simulations, or inside the selected time
window, and construct histograms of return times for different
system parameters. All return times are scaled by the average
return time $R$ in each particular simulation. Therefore, the
second set of statistical measures consists of the mean return
time $R$, the slope of the return time probability function $P_r$,
and moments derived from them.

\section{Results and discussion}

The first step was to generate numerically rational solutions
which can be classified as RWs. The existence of these solutions
had been related, at least initially, with the development of MI
\cite{6,13}, which is by itself threshold determined. Intensive
numerical checking has shown that the rational solutions of the
types presented in \cite{6,13}, see also Figs. \ref{fig1} and
\ref{fig2}, can be obtained from both coherently or noise driven
MI \cite{14}, and represent short-lived or transient wave
structures. It should be noted that the exact choice of the
initial excitation is crucial for the generation of rational
solutions in the defocusing case. In this respect, the structures
which were analytically derived in \cite{13} from eigenvalues of
the Manakov system, have only been observed in the initial phase
of the development of baseband MI development, in the presence of
a periodically perturbed plane wave background, additionally
modulated by a super Gaussian.

Regardless of the initial perturbations, the long term dynamics of
high amplitude events in the Manakov system, observed in the
presence of MI, shows similar tendencies. This is the case for
both types of nonlinearity, that is either focusing or defocusing.
Therefore, statistical ensembles were obtained from long term
numerical simulations involving a noise seeded plane-wave field as
an input condition for the Manakov system (\ref{eq1}). As
discussed in the next section, the width of the calculation window
was adapted in each case in order to include all relevant regimes
of high amplitude events.

\subsection{High amplitude events in the focusing case}

\begin{figure}[h]
\centering
\includegraphics[width=6cm]{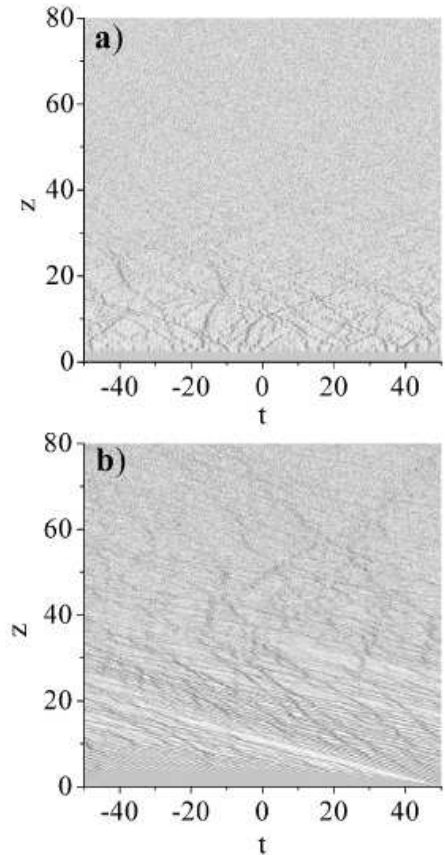}
\caption{Amplitude evolution plots for the focusing Manakov system
with initial plane-wave parameters: (a) $a_1=a_2=1,\,q=0
\,(q_1=q_2=0)$ and (b) $a_1=a_2=1,\,q=1 \,(q_2=-q_1=1)$. Both sets
of parameters belong to the MI development region. Two regions in
these plots can be distinguished with respect to the presence of
isolated localized patterns: the region up to $z \approx 30$ and
above $z \approx 50$, respectively (indicated by dashed lines).
Since the purpose of this figure is to provide a general picture
of the dynamics of the system, we only plot the amplitudes of the
first component of the vector fields, for the sake of simplicity.
Maximum wave amplitude is ~$4.5$ (a) and ~$4$ (b). } \label{fig3}
\end{figure}

The evolution of wave amplitudes for two different initial
conditions, corresponding to parameters above the MI threshold is
presented in Fig.\ref{fig3}. Two different regimes can be
distinguished on these plots: an initial, transient phase, and a
long-term (long propagation lengths) phase. The transient phase is
characterized by the existence of distinct high amplitude
localized patterns, which can be associated with localized bright
and dark rational structures on a finite background. The latter
phase has a highly irregular (turbulent) appearance. These
qualitative differences are reflected on the respective
statistical measures in that they exhibit a different dependence
on the width of the temporal calculation window.

On the basis of numerical simulations, we can distinguish between
an initial, transient, and subsequent long-term dynamical regime
for the ensemble of the high amplitude events. Inside the
transient regime, MI is expected to be the governing mechanism for
the creation of localized waves, including the rational solutions.
These extreme waves appear and disappear, and interact among
themselves and the noisy background upon the propagation. This
behavior relatively quickly evolves into a 'turbulent' like, i.e.
irregularly looking, long-term regime. Interestingly, this kind of
dynamical behavior starts to prevail sooner or later in time,
depending on the specific system parameters, but it is always the
final state of the system. In order to exclude the numerical
uncertainty as a reason for such system behavior, we repeated our
simulations with sympletic variants of the split-step numerical
procedure. Qualitatively, the same results and conclusions were
always obtained.

%%%%%%%%%%%%%%%%%%%%%%%%%%%%%%%%%%%%%%%%%%%%%%%%%%%%%%%%%
%%%%%%%%%%%%%%%%%%%%%%%%%%%%%%%%%%%%%%%%%%%%%%%%%%%%%%%

\begin{figure}[h]
\centering
\includegraphics[width=9cm]{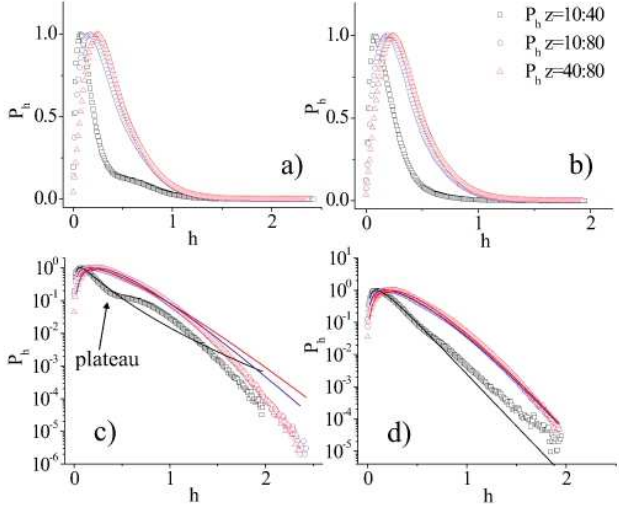}
\caption{The $P_{h}$ vs. $h$ in linear (upper plots) and
semi-logarithmic scales (lower plots) for initial plane-wave
parameters: (a,c) $a_1=a_2=1,\,q=0$ and (b,d) $a_1=a_2=1,\,q=1
\,(q_2=-q_1=1)$. Different curves correspond to the height
distributions of events belonging to different $z$ ranges: black
squares $z=10$ to $40$, blue circles $z=10$ to $80$ , red
triangles $z=40$ to $80$. In plots (c) and (d), solid lines
present GGD fits of the corresponding $P_{h}$ curves. }
\label{fig4}
\end{figure}

Now concerning the statistical measures, the HPD curves (i.e.
$P_h$ vs. wave component height) for the sets of parameters
corresponding to Fig. \ref{fig3} are presented in Fig.
\ref{fig4}((a) and (b)) in a linear scale, and in Fig. \ref{fig4}
((c) and (d)) in a log-linear scale. The statistical distributions
are obtained for different intervals of the evolution coordinate
$z$, as indicated in the legend of Fig. \ref{fig4}. As far as the
overall behavior of these curves is concerned, we may observe that
the $P_h$ curves that characterize the statistics of extreme waves
in the initial phase (black squares) differ from those obtained
for the irregular phase (red triangles). Also, the $P_h$ curves
associated with the entire system dynamics (i.e., both initial and
irregular phases), which are represented by blue circles, almost
coincide with the curves for the long-term phase. One more feature
that is evident, is that the maximum of the $P_h$ curves shifts
towards bigger heights as the calculation window 'moves' in time.
This leads to the conclusion that extreme events occurring in the
later, turbulent phase dominate tail distribution associated with
RW generation.

In addition, we searched for the best fitting function for the
HPDs, following the ideas already presented in the literature \cite{new,new1}.
As expected, the observed HPD deviates from a Gaussian probability distribution (this is a
known feature of RW statistics). Alternatively, it is possible to model the HPD by means of a
generalized Gamma distribution (GGD) \cite{new}. The GGD is often
used in statistics for describing extreme events and it reads as

\begin{equation}
P(x;a,\beta,m)=\frac{a}{\beta\Gamma(m)}\left(\frac{x}{\beta}\right)^{am-1}e^{-(x/\beta)^a},\label{eq7}
\end{equation}

\noindent where $a$ and $m$ are shape parameters, and $\beta$ is a
scale parameter. In order to account for the normalization of our
HPDs, the GGD distribution was multiplied with a parameter $c$,
($0<c<1$). From Fig. \ref{fig4}(c) and Fig. \ref{fig4}(d), it is
obvious that the HPDs associated with the long-term (blue circles)
and the irregular phase (red triangles) are better fitted with the
GGD than the distribution corresponding to the initial phase
(black squares), where the discrepancy is most pronounced in the
tail sections. Also, it is evident that the agreement is better
for the second set of parameters (Fig. \ref{fig4}(d)). However,
although the GGD appears to be the function of choice in the
interaction region, still it does not reduce to any of the special
functions (e.g., Log-normal, Weibull, etc...). The reason for this
could be found in the complexity of the processes governing the
system behavior. The values of the optimal fitting parameters $a$,
$m$ and $\beta$ are listed in Table \ref{tabela1}.
%%%%%%%%%%%%%%%%%%%%%%%%%%%%%%%%%%%%%%%%%%%%%%%%%%%%%%%%%
%%%%%%%%%%%%%%%%%%%%%%%%%%%%%%%%%%%%%%%%%%%%%%%%%%%%%%%

The corresponding values of the significant height, threshold
height and $P_{ee}$ are listed in Table \ref{tabela}.  All of
these quantities were derived from the $P_h$ distribution. The
values of $h_s$ and $h_{th}$ are of the same order in both
selected parameter cases and calculation windows. The values of
$P_{ee}$ in the transient and long-term regimes are similar and
very small, of the order of $0.001$, i.e. $0.1\%$. Depending on
the values of the parameters (amplitudes and phases of initial
plane-wave excitation), and therefore on the the position of the
MI borderline, the value of $P_{ee}$ has a slight tendency to
increase in the transient regime up to $1\%$. The 'plateau'
(indicated by arrow in Fig. \ref{fig4}) in the shape of the
corresponding $P_h$ curves at medium heights, which are observed
in certain parameter areas close to the mentioned border, could be
associated with a zero, or small value of the initial phase
difference between the wave components of the plane wave $q$, see
Fig. \ref{fig4}.

On the other hand, in the presence of nonzero $q$, the
transversely moving localized transient modes can be excited via
the MI mechanism. In addition, for small heights, the growth rate
of $P_h$ with $h$ is larger for simulations involving the
long-term evolution, when compared with the corresponding growth
rate in the early regions where the localized amplitude patterns
are clearly visible. In general, this leads to smaller values of
$P_h$ in the early regime of evolution. Qualitative differences of
the $P_h$ curves corresponding to different calculation windows
undoubtedly show that different types of high amplitude events
govern the system behavior in the course of the vector wave
propagation. On the other hand, the observed negligible
quantitative differences in the $P_{ee}$ indicate the necessity to
search for suitable quantifiers of the types of RWs and their
dynamics. Once again, this opens the question whether the
criterion for RWs based on the significant height is a necessary
and a sufficient one.

\begin{figure}[h]
\centering
\includegraphics[width=9cm]{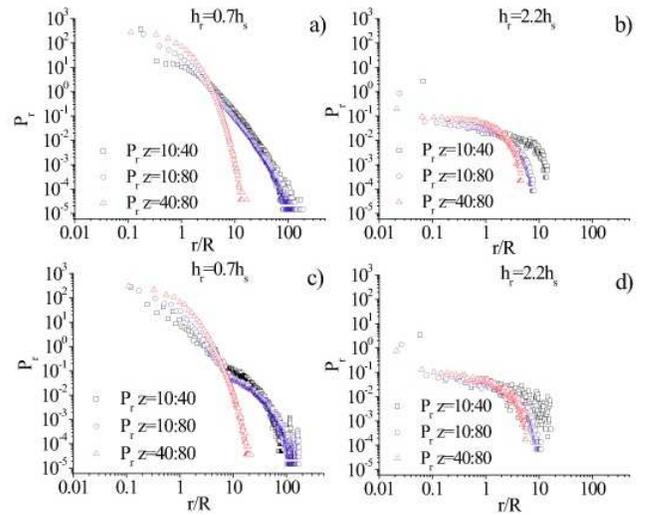}
\caption{The $P_{r}$ vs. $r/R$ for (a), (b) $a_1=a_2=1,\,q=0$ and
(c), (d) $a_1=a_2=1,\,q=1 \,(q_2=-q_1=1)$, with respect to
different threshold amplitudes $h_r$. Three curves in plots (a)
and (c) are obtained for $h_r=0.7 h_s$ and those in (b) and (d)
for $h_r=2.2 h_s$. Different curves correspond to different
calculation windows, as in the previous figure \ref{fig4}. The
amplitude threshold values are summarized in Table II due for
completeness.} \label{fig5}
\end{figure}

An additional set of statistical measures for the RWs was derived
from the statistics of the return time probability, $P_r$, as
shown in Fig. \ref{fig5}. The $P_r$ curves for two different
initial conditions and with respect to (two) different thresholds,
$h_r$, are comparatively presented in this figure. The shape of
the $P_r$ curves changes with the position of the calculation
window and its width, as well as with the amplitude thresholds.
For lower thresholds, the $P_r$ curves corresponding to either
transient or transient+long term evolution phases exhibit a
similar behavior, except for the region corresponding to short
return times, see Fig. \ref{fig5} (a) and (c). The last finding
can be associated with the higher influence of the MI mechanism in
the transient regime, i.e., the short-lived high amplitude
structures are more significant here.

Additionally, for certain initial conditions, one can observe a
turning point, i.e., a plateau, in the region of moderate values
of the return time. By moving the calculation window from the
early transient regime into the long-term limit, the slope of the
$P_r$ curves changes, and it becomes steeper. However, the tails
of all these curves are power-law like. In the long-term regime, a
plateau is no longer present on the $P_r$ curves. All of this
indicates the more frequent appearance of high amplitude events in
the transient phase than in the long term situation. The
distinction between the $P_r$ curves obtained in different
evolution windows is lost with respect to highest amplitude
events, that are associated here with the condition $h_r=2.2 h_s$,
see Fig. \ref{fig5} (b) and (d). In summary, the MI leads to a
transient system behavior, whereas the interactions between moving
'space-time' localized structures become more significant as the
system evolution progresses further. Depending on the system
parameters, the length (i.e. the duration) of the transient phase
will change. Note that the complexity of the dynamics in transient
region, by itself, stems out from the possibility to excite
different types of localized rational or exponentially localized
solutions.

%%%%%%%%%%%%%%%%%%%%%%%%%%%%%%%%%%%%%%%%%%%%%%%%%%%%%%%%%
%%%%%%%%%%%%%%%%%%%%%%%%%%%%%%%%%%%%%%%%%%%%%%%%%%%%%%%

\subsection{High amplitude events in the defocusing case}

The same approach of the previous subsection can also be
applied to study RW statistics in the defocusing Manakov system. The particularity of
this case is the strict dependence of the wave dynamics on the initial conditions, as already mentioned
in Section II. The preparation of initial conditions in our
numerical experiments differs from that presented in \cite{13},
where the presence of baseband MI was declared as a sufficient condition for the
creation of rational/semi-rational solutions. In order to extract
any hidden correlation within our findings, we present the results
for two set of parameters: $a_1=a_2=1, q=0$ and
$a_1=a_2=0.8, q_2=-q_1=1, q=1$, which are outside and inside the
baseband MI region, according to Eq. (\ref{eq6}), respectively. It
should be noted that the properties and values of statistical measures
can strongly depend on the system parameters, which are directly related to the
position of the border of baseband MI, and the value of its growth
rate.

\begin{figure}[h]
\centering
\includegraphics[width=6cm]{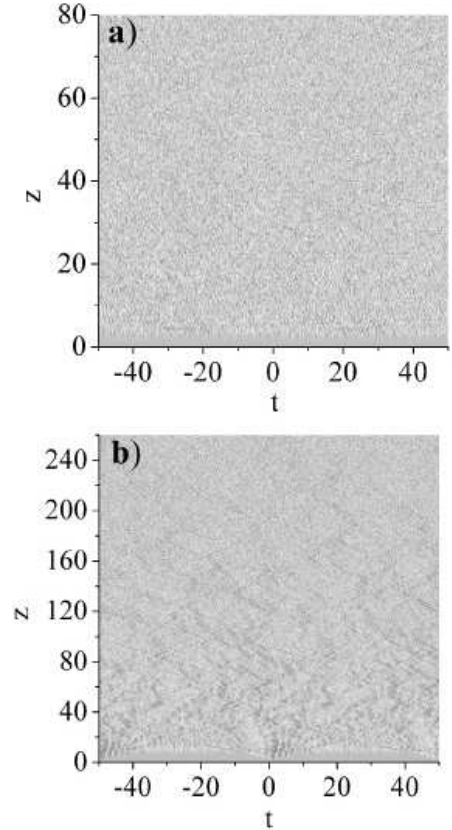}
\caption{Amplitude evolution plots in the defocusing Manakov
system with initial plane-wave parameters: (a) $a_1=a_2=1,\,q=0$
and (b) $a_1=a_2=0.8,\,q=1 \,(q_2=-q_1=1)$. First set of
parameters is outside the baseband MI region, while the second is
inside of it. Maximum amplitude is ~4 (a) and ~3 (b). }
\label{fig6}
\end{figure}

The amplitude plots for both representative parameter sets are
presented in Fig. \ref{fig6}. A clear distinction between two evolution
phases, which was apparent in the focusing case, is absent in
the defocusing regime. However, as we shall see below, the statistical study still shows that, in general,
a competition exists among two different mechanisms for creating the high amplitude events.
Namely, the competition between baseband MI and wave interactions, as well as the prevalence of the
second mechanism in the long-term evolution.

%%%%%%%%%%%%%%%%%%%%%%%%%%%%%%%%%%%%%%%%%%%%%%%%%%%%%%%%%
%%%%%%%%%%%%%%%%%%%%%%%%%%%%%%%%%%%%%%%%%%%%%%%%%%%%%%%

\begin{figure}[h]
\centering
\includegraphics[width=9cm]{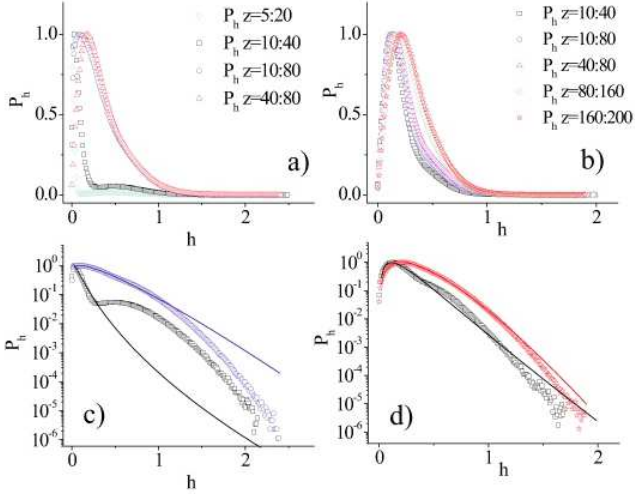}
\caption{The $P_{h}$ vs. $h$ in linear (upper plots) and
semi-logarithmic scales (lower plots) for initial plane-wave
parameters: (a,c) $a_1=a_2=1,\,q=0$ and (b,d) $a_1=a_2=0.8,\,q=1
\,(q_2=-q_1=1)$. Different curves correspond to the height
distributions of events belonging to regions of $z$ are explicitly
reported in the plots. On plots (c) and (d), solid lines represent GGD
fits of corresponding $P_{h}$ curves. } \label{fig7}
\end{figure}

In Fig. \ref{fig7} we present the wave height probability $P_h$
curves together with their GGD fits, for a set of parameters that
are either outside (i.e., $a_1=a_2=1,\,q=0$, see Fig. 7(a) and
(c))) or inside (i.e., $a_1=a_2=0.8,\,q=1$, see Fig. 7(b) and (d))
the range of existence of baseband MI, respectively.  We may note
here the same qualitative behavior for the shape of $P_h$ as
previously observed in the focusing case. Once again, the peak of
the $P_h$ curves shifts towards larger heights as the computation
window progresses to include longer term evolutions. The large
'dip' on the $P_h$ curves in the region of medium $h$ for $q=0$,
i.e., outside of the baseband MI range, is lost in the long-term
calculation windows. A similar 'dip' was previously observed in
the focusing case, where it was associated with the absence of an
initial transverse kick, or phase difference between the
components of the weak initial wave perturbation. In the
defocusing case, this feature can also be related with the absence
of baseband MI \cite{13}.

On the other hand, Fig. \ref{fig7} shows that the $P_h$ behavior
for calculation windows in the long-term range is statistically
the same in both selected parameter cases, namely, either outside
or inside the region for baseband MI. Therefore, based on our
results, one cannot claim that rational solutions, which have been
reported to be a main candidate for RWs in the region of baseband
MI, provide the only source of statistically significant high
amplitude events in the case $a_1=a_2=0.8,\,q=1$ (i.e., inside MI
region). In fig. \ref{fig7} the shapes of $P_h$ curves as well as
the values of $h_s$ (see Table \ref{tabela}) and $P_{ee}$ do not
show a notable dependence upon the size and position of the
calculation window. In general, the values of $h_s$  and $P_{ee}$
follow the same scenario as they did in the focusing case. The
$P_{ee}$ values are very small, of the order $0.1\%-1\%$, in all
parameter regions which are related with the existence of high
amplitude events (rational solitons). Modeling the HPD curves with
the GGD gave similar results as in the focusing case (Fig. 7(c)
and (d)). The agreement between the GGD and HPD is better for the
case of initial parameters inside the baseband MI region,
especially in the long-term limit. Moreover, in this region, one
can notice that the HPDs have a similar shape for both sets of
initial parameters. The values of GGD parameters are given in
Table \ref{tabela1}.

%%%%%%%%%%%%%%%%%%%%%%%%%%%%%%%%%%%%%%%%%%%%%%%%%%%%%%%%%
%%%%%%%%%%%%%%%%%%%%%%%%%%%%%%%%%%%%%%%%%%%%%%%%%%%%%%%

\begin{figure}
\centering
\includegraphics[width=9cm]{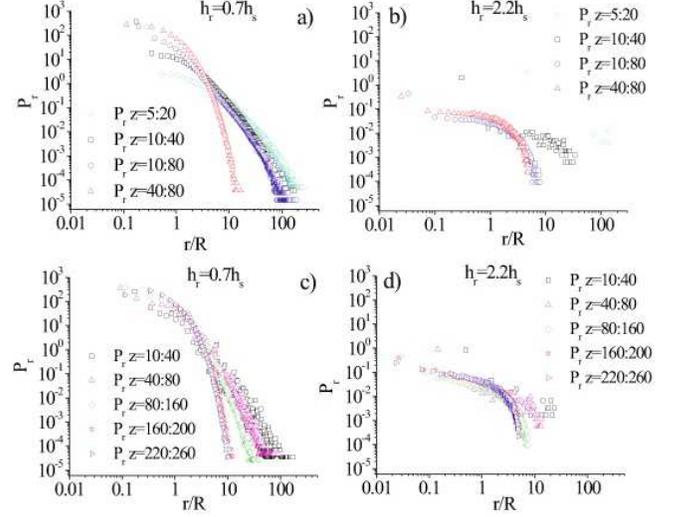}
\caption{The $P_{r}$ vs. $r/R$ for (a, b) $a_1=a_2=1,\,q=0$ and
(c, d) $a_1=a_2=0.8,\,q=1 \,(q_2=-q_1=1)$, with respect to
different threshold amplitudes $h_r$. Three curves in plots (a)
and (c) are obtained for $h_r=0.7 h_s$ and those in (b) and (d)
for $h_r=2.2 h_s$. Different curves correspond to different
calculation windows, as in previous figure \ref{fig7}.}
\label{fig8}
\end{figure}

On the other hand, the return probability $P_r$ behavior is
illustrated in Fig. \ref{fig8}. For higher values of the threshold
amplitude ($h_r=2.2 h_s$), the $P_r$ curves show the same tendency
with respect to the position of the calculation window for both
sets of parameters. By moving the calculation windows towards the
long-term region, the slopes of the $P_r$ curves for lower
threshold ($h_{r}=0.7 h_s$) increase, whereas the shape of the
$P_r$ curves does not change with further changes in the position
or (width) of the calculation window.

A similar tendency regarding the shape of the $P_r$ curves can be
recognized for higher threshold values. By comparing the return
times of high amplitude events for the two selected thresholds, we
can conclude that the return time of the highest amplitude events
is smaller than for the rest of the selected events. This is in
accordance with the values of $R$ which are presented in Table
\ref{tabela}. Note that this is the case for both sets of
parameters, i.e., either outside or inside the baseband MI region.
On the other hand, the differences in $P_r$ and related quantities
for calculations windows in the 'transient' phase are obvious, and
can be related to different types of RWs with respect to those in
the latter phases of the system evolution. In general, for smaller
thresholds, the slopes of the $P_r$ curves change in a way similar
to that observed in the focusing regime. Namely, the slope of the
curve in the long-term (turbulent) regime is steeper than in the
transient phase. This correlates with the mechanism responsible
for exciting high amplitude events. In the first case, the RW
generation is associated with MI, whereas in the second case it is
associated with interactions between different high amplitude
modes. Note that the $P_r$ for both vector field components were
calculated, and we confirmed that they obey the same statistical
scenario.

%%%%%%%%%%%%%%%%%%%%%%%%%%%%%%%%%%%%%%%%%%%%%%%%%%%%%%%%%
%%%%%%%%%%%%%%%%%%%%%%%%%%%%%%%%%%%%%%%%%%%%%%%%%%%%%%%

\begin{table}
\caption{GGD fitting parameters for plots in Figs. \ref{fig4} and
\ref{fig7}. The shape parameter values for certain standard
distributions are: Gamma distribution, $a=1$, exponential
distribution, $a=1,\, m=1$, Rayleigh $a=2,\, m=1$, Weibull
distributions $m=1$, and log-normal distribution $m\rightarrow
\infty$.} \label{tabela1}
\begin{tabular}{|l|l|l|l|} \hline
\ \ Fig. 4a \ \ &\ \ $a$ \ \ &\ \ $m$ \ \ & \ \ $\beta$ \ \
\\ \hline \hline
\ \ $z=10:40$ &  \ \ 0.430 & \ \ 5.926 & \ \ 0.003\\
 \hline
\ \ $z=10:80$ &  \ \ 1.145 & \ \ 1.381 & \ \ 0.294\\ \hline \ \
$z=40:80$ &  \ \ 1.082 & \ \ 2.047 & \ \ 0.211\\
\hline \ \ Fig. 4b \ \ &\ \ $a$ \ \ &\ \ $m$ \ \ & \ \ $\beta$ \ \
\\ \hline \hline
\ \ $z=10:40$ & \ \ 0.908& \ \ 2.060& \ \ 0.088 \\ \hline \ \
$z=10:80$ & \ \ 1.264& \ \ 1.380& \ \ 0.270 \\ \hline \ \
$z=10:40$ & \ \ 1.217& \ \ 1.799& \ \ 0.234 \\ \hline \ \ Fig. 7a
\ \ &\ \ $a$ \ \ &\ \ $m$ \ \ & \ \ $\beta$ \ \
\\ \hline \hline
\ \ $z=5:20$ & \ \ 1.246& \ \ 1.148& \ \ 0.019 \\ \hline \ \
$z=10:40$ & \ \ 0.443& \ \ 5.473& \ \ 0.002 \\ \hline \ \
$z=10:80$ & \ \ 1.163& \ \ 1.077& \ \ 0.343 \\ \hline \ \
$z=40:80$ & \ \ 0.741& \ \ 3.234& \ \ 0.072 \\ \hline \ \ Fig. 7b
\ \ &\ \ $a$ \ \ &\ \ $m$ \ \ & \ \ $\beta$ \ \
\\ \hline \hline
\ \ $z=10:40$ & \ \ 0.770& \ \ 3.420& \ \ 0.041 \\ \hline \ \
$z=10:80$ & \ \ 0.715& \ \ 3.765& \ \ 0.037 \\ \hline \ \
$z=40:80$ & \ \ 0.735& \ \ 3.633& \ \ 0.044 \\ \hline \ \
$z=80:160$ & \ \ 1.143& \ \ 1.924& \ \ 0.179 \\ \hline \ \
$z=160:200$ & \ \ 1.278& \ \ 1.657& \ \ 0.229 \\ \hline
\end{tabular}
\end{table}

\begin{table}
\caption{Values of different parameters derived from the $P_h$ and
$P_r$ for focusing and defocusing cases. } \label{tabela}
\begin{tabular}{ |p{1.5cm}|p{1,7cm}|p{1.7cm}|p{1.7cm}|  }
\hline
\multicolumn{4}{|c|}{$a_1=a_2=1,\,q=0$} \\
\hline
focussing& $z=[10:40]$ &$[10:80]$& $[40:80]$\\
\hline
$h_s$& $0.549$ &$0.673$ &$0.697$\\
$h_{th}$ & $1.210$ &$1.480$&$1.530$\\
$P_{ee}$&$0.006$ &$0.013$ &$9.96e^{-4}$\\
$R_1$& $0.011$ &$0.009$&$0.009$\\
$R_2$ & $0.026$ &$0.021$ &$0.019$\\
$R_3$& $2.019$ &$8.279$ &$8.399$ \\
\hline \hline
\multicolumn{4}{|c|}{$a_1=a_2=1,\,q=1$} \\
\hline
focussing& $z=[10:40]$ &$[10:80]$& $[40:80]$\\
\hline
$h_s$& $0.397$ &$0.592$ &$0.631$\\
$h_{th}$ & $0.836$ &$1.302$&$1.349$\\
$P_{ee}$&$0.008$ &$0.002$ &$0.002$\\
$R_1$& $0.015$ &$0.011$&$0.009$\\
$R_2$ & $0.039$ &$0.024$ &$0.019$\\
$R_3$& $2.803$ &$4.566$ &$6.600$ \\
\hline
\end{tabular}

\begin{tabular}{ |p{1cm}|p{1.7cm}|p{1.7cm}|p{1.7cm}|  }
\hline
\multicolumn{4}{|c|}{$a_1=a_2=1,\,q=0$} \\
\hline
defoc.& $z=[10:40]$ &$[10:80]$& $[40:80]$\\
\hline
$h_s$& $0.506$ &$0.651$ &$0.666$\\
$h_{th}$ & $1.110$ &$1.430$&$1.465$\\
$P_{ee}$&$0.011$ &$0.002$ &$0.001$\\
$R_1$& $0.011$ &$0.010$&$0.009$\\
$R_2$ & $0.027$ &$0.023$ &$0.021$\\
$R_3$& $0.744$ &$8.292$ &$8.068$ \\
\hline
\end{tabular}

\begin{tabular}{ |p{0.7cm}|p{1.5cm}|p{1.3cm}|p{1.3cm}| p{1.3cm}| p{1.4cm}|}
\hline
\multicolumn{6}{|c|}{$a_1=a_2=0.8,\,q=1$} \\
\hline
defoc.& $z=[10:40]$ &$[10:80]$& $[40:80]$&$[80:160]$ &$[160:200]$\\
\hline
$h_s$& $0.411$ &$0.466$ &$0.498$&$0.534$&$0.553$ \\
$h_{th}$ & $0.904$ &$1.025$&$1.096$&$1.175$&$1.217$ \\
$P_{ee}$&$0.003$ &$0.002$ &$0.002$&$9.6e^{-4}$&$7.7e^{-4}$ \\
$R_1$& $0.011$ &$0.011$&$0.011$&$0.009$&$0.009$ \\
$R_2$ & $0.023$ &$0.023$ &$0.023$&$0.018$&$0.018$ \\
$R_3$& $1.013$ &$4.221$ &$2.779$ &$10.444$&$7.663$ \\
\hline
\end{tabular}

\end{table}

\section{Conclusions}

Let us summarize the results of our study of high amplitude events
in the Manakov system, by pointing out the main findings. In both the focusing and
defocusing nonlinearity regime, it was shown that the type of
initial perturbation of the plane wave background did not have a
significant influence on the long term evolution of high amplitude
events. On the other hand, we found that the properties of the long term evolution can be associated with the
presence of MI in the Manakov system. In
latter stages of the evolution of the high amplitude modes, their
interactions drive the dynamics of high amplitude events, and potentially affect the
properties and the behavior of RWs. This conclusion does not
depend on the character of nonlinearity in the Manakov system.

We decided to use the term 'high amplitude events' instead of
'RWs', on the basis of the unclear indications about the criteria for extracting RWs from
a statistical analysis based on the height and return time
probabilities. We have found that the statistics of heights of
high amplitude events can be described very well by the generalized Gamma distribution in
both the focusing and the defocusing regime, especially in the long term
propagation limit, i.e., in a regime where  interactions between
the high amplitude events are shown to be the most prominent
contributors to the RW generation. In contrast, we have shown that,
in order to identify a high amplitude event as a RW, new
criteria are necessary, at least in multi-component systems.
The significant height vector equivalent of the corresponding
scalar quantity is not sensitive enough, in order to clearly identify the
RW, as well as to distinguish between different types of RWs.

Data derived from the return time probability mostly confirm
previous statements, and show that the return time based
quantities can be promising candidates for good classifiers of
different types of RWs. The significance of the initial system
preparation, width and position of the calculation window, on the
values of the threshold amplitudes has been pointed out.
Therefore, the main contribution of this study is the suggestion
and development of a new strategy for confirming the basic
properties of different RWs events in multicomponent nonlinear
wave systems.

\acknowledgments A. M., Lj. H., and A. M. acknowledge support from
Ministry of Education, Science, and Technological Development of
Republic of Serbia [III 45010]. This project was partially
supported from the European Union's Horizon 2020 research and
innovation programme under the Marie Sklodowska-Curie grant
agreement No 691051. The work of S.W. was supported by the Russian
Ministry of Science and Education (Grant 14.Y26.31.0017).

\end{document}